\documentclass [11pt,a4paper] {article}

\usepackage[cp1252]{inputenc}
\usepackage[english]{babel}

\usepackage{amssymb}
\usepackage{amsmath}
\usepackage{amsfonts,amssymb}

\usepackage[dvips]{graphicx}

\begin{document}

\title{Computational Solution to Quantum Foundational Problems}

\author{Arkady Bolotin\footnote{$Email: arkadyv@bgu.ac.il$} \\ \textit{Ben-Gurion University of the Negev, Beersheba (Israel)}}

\maketitle

\begin{abstract}\noindent This paper argues that the requirement of applicableness of quantum linearity to any physical level from molecules and atoms to the level of macroscopic extensional world, which leads to a main foundational problem in quantum theory referred to as the ``measurement problem'', actually has a computational character: It implies that there is a generic algorithm, which guarantees exact solutions to the Schrödinger equation for every physical system in a reasonable amount of time regardless of how many constituent microscopic particles it comprises. From the point of view of computational complexity theory, this requirement is equivalent to the assumption that the computational complexity classes \textbf{P} and \textbf{NP} are equal, which is widely believed to be very unlikely. As demonstrated in the paper, accepting the different computational assumption called the Exponential Time Hypothesis (that involves $\bf{P}\!\!\neq\!\bf{NP}$) would justify the separation between a microscopic quantum system and a macroscopic apparatus (usually called the Heisenberg cut) since this hypothesis, if true, would imply that deterministic quantum and classical descriptions are impossible to overlap in order to obtain a rigorous derivation of complete properties of macroscopic objects from their microstates.\\

\end{abstract}

\section{Introduction}

\noindent In quantum theory, the state of a microscopic particle is described by a state-vector $\!\left.\left|\psi \!\left(t\right)\!\right.\right\rangle$ (identified with a ray in the Hilbert space of the particle), whose subsequent time evolution is governed by the Schrödinger equation: $i\hbar\:{\partial\!\!\left.\left|\psi \!\left(t\right)\!\right.\right\rangle}/{\partial t}=H_p\!\left.\left|\psi \!\left(t\right)\!\right.\right\rangle$, where $H_p$ is the Hamiltonian of the particle. According to the Copenhagen interpretation of quantum mechanics, the Schrödinger equation applies when the microscopic particle is evolving in isolation. But if the particle interacts with a macroscopic system or a measuring apparatus (like, for example, in the Stern--Gerlach setup), the reduction postulate and the Born rule must be used.\\

\noindent But then again, it seems unreasonable to have two incompatible dynamical laws governing the time evolution of the same particle: the deterministic Schrödinger equation for the events at the microscopic level and the stochastic reduction postulate for the events associated with micro--macro interactions. More logical and -- in accordance with Occam's razor \cite{Soklakov} -- more intellectually economical is to believe that the Schrödinger equation is applicable to the whole class of physical entities and hence governs all phenomena at both the microscopic and macroscopic levels (in fact, this is a belief accepted at the present by the great majority of practicing physicists \cite{Schlosshauer04}).\\

\noindent Consequently, the time evolution of the state-vector $\!\left.\left|\Psi\!\left(t\right)\!\right.\right\rangle $ describing the state of the macroscopic composite system comprised of the microscopic particle and the apparatus may also be defined by the Schrödinger equation $i\hbar\:{\partial\!\left.\left|\Psi\!\left(t\right)\!\right.\right\rangle}/{\partial t}\!=\!H\!\left.\left|\Psi\!\left(t\right)\!\right.\right\rangle$, so that knowing the state of the composite system at an initial time $t\!=\!T_{\! i}$, one can in principle solve the Schrödinger equation with this initial condition to predict the state of the system at any future time $t\!=\!T_{\! f}$. As the Schrödinger Hamiltonian $H$ is Hermitian, the predicted state at the time $T_{\! f}$ is related to that at the initial time $T_{\! i}$ by the deterministic relation $\!\left.\left|\Psi\!\left(T_{\! f}\right)\!\right.\right\rangle\!=\!U\!\left(T_{\! f},T_{\! i}\right)\!\!\left.\left|\Psi\!\left(T_i\right)\!\right.\right\rangle$, where the transition operator $U\!\left(T_{\! f},T_{\! i}\right)$ is unitary and completely specified by the Hamiltonian $H$ of the composite system: $U\!\left(T_{\! f},T_{\! i}\right)={\exp \!\left(-i\!\left(T_{\! f}\!-\!T_{\! i}\!\right)\!H\!\!/\hbar \!\right)}$.\\

\noindent However, as soon as the prediction $\!\left.\left|\Psi\!\left(T_{\! f}\right)\!\right.\right\rangle$ is extracted from the Schrödinger equation, one will get -- in virtue of linearity of this equation -- a superposition state of the particle plus the apparatus (the infamous Schrödinger cat state), which is never experienced in our classical world. This paradox is especially puzzling since apparently the Schrödinger equation contains nothing prohibiting its application to macroscopic objects. Particularly, this equation does not say how large objects must be, before they can be said to obey the equations of Newtonian mechanics. Thence, a belief in the generic applicableness of the Schrödinger equation underlies a main foundational problem in quantum theory (known by the different names such as \textit{macro-objectification problem}, \textit{problem of definite outcomes}, \textit{quantum measurement problem}, \textit{problem of the emergence of classicality from quantum systems}, and so on).\\

\noindent A related problem also stemming from the belief in the generic applicableness of the Schrödinger equation is the origin of the probabilities in quantum theory. Namely, how can the probabilities come out of a deterministic, continuous and unitary time evolution given by the Schrödinger equation?\\

\noindent Those foundational problems have been debated for over eighty years, and during all these years, most of effort went into trying to resolve the problems within a physical theory (or physical theories). Thus, various attempts were made to change the interpretational rules of quantum theory as well as to modify its foundations (see papers \cite{Styer,Bassi,Laloe} to name a few) including attempts \cite{Penrose} to use the principles of general relativity to change the formalism of quantum mechanics. But since none of these attempts has gained general acceptance it might be time to look beyond physics and try to resolve those quantum foundational problems within a mathematical theory, specifically, computational complexity theory.\\

\noindent Indeed, along with the comprehensible and explicit assumption of the universality of the physical laws, the belief in the generic applicableness of the Schrödinger equation contains an additional, implicit assumption that has rather an \textit{algorithmic} (or computational) character. This additional assumption is that there is a way to \textit{efficiently} extract (i.e., extract \textit{in a reasonable amount of time}) predictions about future states of physical systems -- microscopic and macroscopic alike -- from the Schrödinger equation. In other words, the hypothesis that the Schrödinger equation is applicable to everything from microscopic particles to macroscopic objects to the whole universe implies that there is a generic algorithm, which guarantees the exact and efficient solution to the Schrödinger equation for every physical system, no matter how complicated and huge it is (i.e., how many constituent microscopic particles it comprises). But what if such an efficient generic algorithm does not really exist? If it were so, then quantum theoretical constructions like ``a quantum state of a macroscopic object'' or ``the wave function of the universe'' would be nothing more than nontestable empty abstractions. Undeniably, this would have a crucial implication for the resolution of the quantum foundational problems.\\

\noindent The aim of this paper is to critically evaluate the assumption of the across-the-board efficient solvability of the Schrödinger equation in order to judge how plausible -- from the point of view of computational complexity theory -- this assumption is. The paper is structured as follows. First, considering the Schrödinger equation as a computational problem, it will be proved that this problem is \textbf{NP}-hard, which means that given a generic algorithm that solves exactly Schrödinger's equation for all possible Hamiltonians, one would be able to solve all problems in the \textbf{NP} complexity class (covering most natural computer science problems). Next, it will be demonstrated that unless the Exponential Time Hypothesis were proved to be false, Schrödinger's equation would be merely unsolvable for macroscopic systems and accordingly inapplicable to their time evolution portrayal. Finally, it will be shown that randomness is entered in pure Hamiltonian evolution as a way to obtain the prediction about the state of a microscopic system interacting with its macroscopic environment without solving the environmental Schrödinger equation -- an intractable computational problem.\\

\noindent Before proceeding with our plan, we should note the following. Since our second goal is to elucidate the origin of probabilities in quantum theory, no new classes (such as \textbf{BQP} and \textbf{QMA}) extending the classical complexity classes \textbf{P} and \textbf{NP} will be considered in the paper. Obviously, allowing for those classes (of decision problems solvable \textit{probabilistically}) for the investigation into the probability origin would be subject to the charge of circularity.\\

\section{Schrödinger's equation as a computational problem}

\noindent Let us start our evaluation by defining the Schrödinger equation as a computational problem, which we will call the problem ${\Phi }_{\Psi }$:

\begin{quotation}
\noindent\textit{\textbf{Problem} ${\mathbf {{\Phi }_{\!\Psi}} }$ Given the Schrödinger Hamiltonian} $\!H\!$ \textit{of a physical system, what is the exact solution }$\!\left.\left|\psi\!\left(t\right)\!\right.\right\rangle$
\textit{to the Schrödinger equation }
$i\hbar\:{\partial\!\left.\left|\psi\!\left(t\right)\!\right.\right\rangle}/{\partial t}\!=\!H\!\!\left.\left|\psi\!\left(t\right)\!\right.\right\rangle$\textit{?}
\end{quotation}

\noindent Despite the fact that in this form the differential operator ${\partial }\!/\!{\partial t}$ and the Hamiltonian $\!H\!$ are just abstract operators acting on kets $\!\left.\left|\psi \!\left(t\right)\!\right.\right\rangle$, abstract objects, the problem ${\Phi }_{\Psi }$ may be understood as \textit{a functional computational problem} once this form is projected into the position basis $\{\!\left.\left|{\mathbf r}\!\right.\right\rangle\}$: $i\hbar\:{\partial\Psi\!\left({\mathbf r},{m},t\right)}/{\partial t}\!=\!H\Psi\!\left({\mathbf r},m,t\right)$, where the wave function $\Psi\!\left({\mathbf r},m,t\right)$ is the scalar product $\!\left\langle {\mathbf r}\left|\psi \!\left(t\right)\!\right.\right\rangle$, ${\mathbf r}=\left({{\mathbf r}}_1,\dots ,{{\mathbf r}}_j,\dots ,{{\mathbf r}}_N\right)$ denotes the sets of position vectors, $m=\left({m}_1,\dots ,{m}_j,\dots ,{m}_N\right)$ signifies the set of discrete variables (spin components along the $z$-axis), each of which $m\!{_j}$ is out of the following set of values (determined by the spin quantum number $s\!{_j}$ of the $j^{\mathrm {th}}$ system constituent particle)

\begin{equation} \label{1} 
   m\!{_j}\in
   \left\{\!
      -s\!{_j}\hbar, -\!\left(s\!{_j}-1\right)\!\hbar, \dots , +\!\left(s\!{_j}-1\right)\!\hbar,  +s\!{_j}\hbar
   \right\}
\;\;\;\;  ,
\end{equation}
\smallskip

\noindent and $N$ is the system constituent particle number. In this way, to solve exactly an instance of the functional problem ${\Phi}_{\Psi}$ would mean to substitute the indeterminate $H$ by the Hamiltonian for a particular system accounting for the kinetic and potential energy of $N$ particles constituting the system in the Schrödinger equation and then to solve exactly the resulting partial differential (in general, time-dependent) equation for the state of the system at time $t$ represented by the vector $\!\left.\left|\psi \!\left(t\right)\!\right.\right\rangle$, which stands for the wave function $\Psi\!\left({\mathbf r},m,t\right)$.\\

\noindent At this point, we take a slight detour and talk about the exact solutions to Schrödinger's equation, which will be central to this paper. Following the papers \cite{Sasaki} and \cite{Zhang}, the Schrödinger equation of a system is exactly solvable if all the eigenvalues and the corresponding eigenfunctions of the system can be determined exactly. In contrast, a system is quasi-exactly solvable if only a finite number of exact eigenvalues and eigenfunctions can be obtained. An immediate consequence of such a characterization of exact solvability for the Schrödinger Hamiltonian $H$ is that it can be diagonalized algebraically and exact, closed-form expressions of the corresponding spectra can be evaluated in a finite number of standard operations. In the case of quasi-exact solvability, the remaining part of the spectrum is not analytically accessible and can only be computed as an approximation (though capable of evaluating in a finite number of standard operations). Seeing that in both cases (quasi and not) an evaluation algorithm terminates after a finite amount of time (instead of looping indefinitely), for the purpose of this paper we will not distinguish between exact and quasi-exact solvability of the Schrödinger equation.\\

\noindent As every function computational problem can be turned into \textit{a decision computational problem} (i.e., a question with only `yes'-or-`no' answer), we can easily change the functional problem ${\Phi }_{\Psi }$ into the decision problem ${\Pi }_{\Psi }$ by incorporating a set of additional restraints $R$ imposed on the solution $\!\left.\left|\psi \!\left(t\right)\!\right.\right\rangle$ (or its eigenvalue) or/and on a set of polynomially bounded functions of the solution $\!\left.\left|\psi \!\left(t\right)\!\right.\right\rangle$ (or its eigenvalue) into the problem ${\Phi }_{\Psi }$:

\begin{quotation}
\noindent\textit{\textbf{Problem} ${\mathbf {{\Pi }_{\Psi }} }$ Given the Schrödinger Hamiltonian} $H$ \textit{of a physical system and a set of restraints} $R$\textit{, does the system have the exact solution }$\!\left.\left|\psi \!\left(t\right)\!\right.\right\rangle$\textit{ to the Schrödinger equation }$i\hbar\:{\partial\!\left.\left|\psi\!\left(t\right)\!\right.\right\rangle}/{\partial t}\!=\!H\!\!\left.\left|\psi\!\left(t\right)\!\right.\right\rangle$\textit{, which is subject to those restrains }$R$\textit{?}
\end{quotation}

\noindent The decision problem ${\Pi }_{\Psi }$ can equally be defined as the set $S_{\Psi }$ of inputs -- various Hamiltonians $H$ of different `sizes', i.e., numbers $N$ of constituent particles, and corresponding restraints\textit{ }$R$ -- for which ${\Pi }_{\Psi }$ returns 1 (i.e. `yes'):

\begin{equation} \label{2} 
   S_{\Psi }=\left\{ 
      \begin{array}{cl}
         H,\ R:
         &
         {\rm \ }{\Pi }_{\Psi }\!\left(i\hbar\:\displaystyle \frac{\partial }{\partial t}\!\left.\left|\psi \!\left(t\right)\!\right.\right\rangle = H\!\!\left.\left|\psi \!\left(t\right)\!\right.\right\rangle \ \wedge \ R\!\right)=1
      \end{array}
   \right\}
\;\;\;\;  .
\end{equation}
\smallskip

\noindent Let $A\!\left({\Phi }_{\Psi }\!\right)$ denote an exact generic algorithm capable of solving exactly the Schrödinger equation for an arbitrary physical Hamiltonian $H$. From this notion it immediately follows that the algorithm $A\!\left({\Phi }_{\Psi }\!\right)$ can exactly solve the functional problem ${\Phi }_{\Psi }$ on all instances, i.e., for all possible physical Hamiltonians $H$ of all possible `sizes'. (One of such exact generic algorithms is well known -- it is \textit{brute force}.) Evidently this algorithm $A\!\left({\Phi }_{\Psi }\!\right)$ can be easily modified to solve all instances of the decision problem ${\Pi }_{\Psi }$ as well: The algorithm for solving ${\Pi }_{\Psi }$ will first call the algorithm $A\!\left({\Phi }_{\Psi }\!\right)$ as a subroutine to solve the Schrödinger equation for the state $\!\left.\left|\psi \!\left(t\right)\!\right.\right\rangle$ and then decide in polynomial number of steps whether the ensuing state $\!\left.\left|\psi \!\left(t\right)\!\right.\right\rangle$ (or its eigenvalue) satisfies the restraints $R$.\\

\noindent Suppose the vector $\!\left.\left|\psi \!\left(t\right)\!\right.\right\rangle$ is the exact solution to the Schrödinger equation for the given Hamiltonian $H$ and subject to the given restrains $R$. Let us show that the decision problem ${\Pi }_{\Psi }$ can be quickly verified, i.e., one can check whether ${\Pi }_{\Psi }$ returns 1 for these $H$ and $R$ in polynomial time. To accomplish this, one should substitute the exact solution $\!\left.\left|\psi \!\left(t\right)\!\right.\right\rangle$ back into the expression for ${\Pi }_{\Psi }$ and estimate the runtime complexity of the operations needed to prove that ${\Pi }_{\Psi }$ indeed returns 1.\\

\noindent Let $L$ be the minimal number of elementary operations sufficient to compute the effects of the differential operator ${\partial }\!/\!{\partial t}$ and the Hamiltonian $H$ on the given vector $\!\left.\left|\psi \!\left(t\right)\!\right.\right\rangle$; we will call $L$ the complexity of verification. The easiest method of approximating partial derivative ${\partial F}\!/{\partial q_j}$ is the finite difference quotient

\begin{equation} \label{3} 
   \frac{\partial F}{\partial q_j}
   =
   \frac{F\left(q_1, \dots ,q_j+h_j,\dots, q_n\right)
   -
   F\left(q_1, \dots ,q_j,\dots, q_n\right)}{h_{\! j}}
\end{equation}
\smallskip

\noindent with $h_{\! j}\to 0$. Hence, using the results of the papers \cite{Paterson,Baur}, in the basis $\{\!\left.\left|{\mathbf r}\!\right.\right\rangle\}$ the complexity $L$ can be presented as follows:

\begin{equation} \label{4} 
   L\left(
        \frac{\partial \!\left.\left|\psi \!\left(t\right)\!\right.\right\rangle}{\partial t},\ H\!\left.\left|\psi \!\left(t\right)\!\right.\right\rangle 
   \right)
   =
   L\left(
         \frac{{\partial }\Psi }{\partial t},\frac{{\partial }^2\Psi }{\partial {{\mathbf r}}^2_1},\dots ,\frac{{\partial }^2\Psi }{\partial {{\mathbf r}}^2_N},\frac{{\partial }\Psi }{\partial {{\mathbf r}}_1},\dots ,\frac{{\partial }\Psi }{\partial {{\mathbf r}}_N}
   \right)
   \le
   O\!\left(N^2\right)
   \!\cdot\!
   {\mit cost}\!\left(\!\Psi\!\right)
\;\;\;\;  ,
\end{equation}
\smallskip

\noindent where only nonscalar arithmetic operations (i.e., binary operations whose both operands involve the function $\Psi$) are considered contributed to the complexity of verification $L$, whereas additions/subtractions and multiplications by arbitrary scalars are allowed for free,  ${\mit cost}\!\left(\!\Psi\!\right)$ denotes the computational cost of the wave function evaluation at particular numerical values ${\mathbf r}$, $m$ and $t$. In order to the interpretation of solutions to Schrödinger's equation to make sense, it must be feasible to evaluate the given exact solution $\Psi$ at any allowable ${\mathbf r}$, $m$ and $t$ because otherwise it would be impossible to use $\Psi\!\left({\mathbf r},m,t\right)$ to compute a measurable observable of the quantum system. In conformity with Cobham's thesis \cite{Cobham}, the function $\Psi\!\left({\mathbf r},m,t\right)$ can be feasibly evaluated only if it can be evaluated on some computational device in polynomial time. This implies that  ${\mit cost}\!\left(\!\Psi\!\right)={\mit poly}\!\left(\!N\!\right)$, and hence, the verification complexity $L$ is upper-bounded by a polynomial. The corollary to this conclusion is that the decision problem ${\Pi }_{\Psi }$ is in the \textbf{NP} complexity class (of computational problems, whose solutions can be verified in polynomial time).\\

\noindent Let $H\!\left({\sigma}_1,\dots ,{\sigma}_j,\dots ,{\sigma}_N\!\right)$ be the Hamiltonian that describe the energy of configuration of a set of $N$ spins $\sigma\!{_j}\hbar=2m\!{_j}\in\left\{\!-\hbar, +\hbar\right\}$ in classical Ising models of a spin glass \cite{Fischer,Guerra}

\begin{equation} \label{5} 
   H\!\left({\sigma}_1,\dots ,{\sigma}_j,\dots ,{\sigma}_N\!\right) 
   =
  -\sum_{j<k}{J_{jk}\sigma\!{_j}\sigma\!{_k}}
  -\mu\sum^N_{j}{h_{j}\sigma\!{_j}}
\;\;\;\; 
\end{equation}
\smallskip

\noindent such that real numbers $J_{jk}$ be coupling (interaction) coefficients, $h\!{_j}$ be external magnetic fields, and $\mu$ be the magnetic moment. Since the generic algorithm $A\!\left({\Phi }_{\Psi }\!\right)$ can solve exactly the Schrödinger equation for all Hamiltonians, it can also solve the Schrödinger equation for the quantum version of the Ising Hamiltonian $H({\sigma}^z_1,\dots ,{\sigma}^z_j,\dots ,{\sigma}^z_N)$, where spins $\sigma\!{_j}$ are simply replaced by quantum operators – Pauli spin-1/2 matrices ${\sigma}^z_j$. On the other hand, the paper \cite{Lucas} explains how ``all the famous \textbf{NP} problems" (such as Karp's 21 \textbf{NP}-complete problems \cite{Karp,Garey}) can be written down as Ising models with only a polynomial number of steps (to be exact, with a polynomial number of spins which scales no faster than $N^3$). Therefore, in just a polynomial number of steps one can get from any \textbf{NP}-complete problem to the Hamiltonian of an Ising spin glass $H\!\left({\sigma}_1,\dots ,{\sigma}_j,\dots ,{\sigma}_N\!\right)$, whose decision problem ${\Pi }_0$ (the family of instances of the problem ${\Pi }_{\Psi }$)

\begin{quotation}
\noindent\textit{\textbf{Problem} ${\mathbf {{\Pi }_0} }$ Given the Ising Hamiltonian} $H\!\left({\sigma}_1,\dots ,{\sigma}_j,\dots ,{\sigma}_N\!\right),$ \textit{does a spin glass have the ground state} $\left.\left|\psi_0\!\right.\right\rangle$ \textit{– i.e., the exact solution to the time-independent Schrödinger equation} $H({\sigma}^z_1,\dots ,{\sigma}^z_j,\dots ,{\sigma}^z_N)\!\left.\left|\psi_0\!\right.\right\rangle\!=\!E_0\!\left.\left|\psi_0\!\right.\right\rangle$ \textit{– whose energy} $E_0\le0?$
\end{quotation}

\noindent solves the \textbf{NP}-complete problem of interest using the generic algorithm $A\!\left({\Phi }_{\Psi }\!\right)$. Consequently, we get to the following conclusion: As an arbitrary \textbf{NP} problem is polynomial-time reducible to any \textbf{NP}-complete problem and thus to the decision problem ${\Pi }_0$ of the Ising model, any problem in \textbf{NP} can be solved exactly by the generic algorithm $A\!\left({\Phi }_{\Psi }\!\right)$ with only polynomially more work.\\

\noindent This conclusion means that the problem ${\Phi }_{\Psi }$ (of solving exactly Schrödinger's equation for an arbitrary physical system) is \textbf{NP}-hard, i.e., at least as hard as (or harder than) any problem in the \textbf{NP} class. A consequence of the \textbf{NP}-hardness of ${\Phi }_{\Psi }$ is that the class\textbf{ NP} would be equal to the class \textbf{P} (of computational problems solvable in polynomial time) if $A\!\left({\Phi }_{\Psi }\!\right)$ were efficient. But if $\bf{P}\!\!\neq\!\bf{NP}$, the generic algorithm $A\!\left({\Phi }_{\Psi }\!\right)$ could not be efficient, i.e., its runtime complexity would not be polynomial in $N$.\\

\noindent Although whether \textbf{P} is equal to\textbf{ NP} or not is a major unresolved problem in computer science, the assumption of the across-the-board efficient solvability of the Schrödinger equation embraces the attitude that $\bf{P}\!\!=\!\bf{NP}$, contrary to widespread (among computational theory scientists) belief that $\bf{P}\!\!\neq\!\bf{NP}$ \cite{Gasarch}.\\

\section{How Schrödinger's cat is explained by computational complexity theory}

\noindent As an entity of classical realm, a truly macroscopic object contains a large (and essentially unchecked) number $N$ of constituent microscopic particles that interact constantly not only with each other but additionally with a great (and also essentially unchecked) number of different physical systems (varying in their properties and scales) within the vast causal horizon for the object. As such, the Hamiltonian of the truly macroscopic object must be a sum of all possible physical Hamiltonians of all possible ‘sizes’, and hence an algorithm capable of solving exactly the Schrödinger equation for this object must be the exact generic algorithm $A\!\left({\Phi }_{\Psi }\!\right)$. Therefore, if \textbf{P}$\neq$\textbf{NP}, the Schrödinger equation for the truly macroscopic object could not be solved exactly in polynomial time.\\

\noindent Nevertheless, even if the conjecture \textbf{P}$\neq$\textbf{NP} is true, \textbf{NP}-hardness of solving exactly the Schrödinger equation for the truly macroscopic object will not give us any information on what kind of super-polynomial running time is possible for the exact generic algorithm $A\!\left({\Phi }_{\Psi }\!\right)$. Particularly, is it possible that this algorithm is \textit{significantly faster} than brute force, which -- according to postulates of quantum mechanics (namely, the postulate that the Hilbert space ${\mathcal H}$ for the composite system containing two subsystems is the tensor product ${\mathcal H}={\mathcal H}_1\!\otimes{\mathcal H}_2$ of the Hilbert spaces ${\mathcal H}_1$ and ${\mathcal H}_2$ for two constituent subsystems) -- runs exponentially in $N$ as $O\!\left(\!2^N\!\right)$?\\

\noindent Suppose that $A\!\left({\Phi }_{\Psi }\!\right)$ is a sub-exponential time algorithm. Since any \textbf{NP}-complete problem – including the 3-SAT problem – can be written down as the decision problem ${\Pi }_0$ of the Schrödinger equation for the quantum Ising Hamiltonian $H({\sigma}^z_1,\dots ,{\sigma}^z_j,\dots ,{\sigma}^z_N)$, it follows that the algorithm $A\!\left({\Phi }_{\Psi }\!\right)$ can solve any \textbf{NP}-complete problem in sub-exponential time. However, according to the widely believed conjecture called \textit{the Exponential Time Hypothesis}, the 3-SAT problem does not have a sub-exponential time algorithm \cite{Impagliazzo,Woeginger,Lokshtanov}. Hence, if the runtime complexity of $A\!\left({\Phi }_{\Psi }\!\right)$ were sub-exponential in $N$, then the Exponential Time Hypothesis could be shown to be false, which would have dramatic implications for many search problems (such as graph $k$-colorability and maximum cliques) as well as for satisfiability algorithms. Thus, most likely, the exact generic algorithm $A\!\left({\Phi }_{\Psi }\!\right)$ solving exactly Schrödinger's equation for a truly macroscopic object could not be significantly faster than brute force.\\

\noindent But can advances in CPU and other technology make the (presumably) exponential-time algorithm $A\!\left({\Phi }_{\Psi }\!\right)$ efficient at least in practical terms? That is, can they help to solve exactly Schrödinger's equation for a truly macroscopic object by brute force in some reasonable amount of time, say, in one year?\\

\noindent To brute force the solution to the Schrödinger equation for a truly macroscopic object in one year will require a computational device to execute each operation (out of roughly ${\rm 2}^{N_{\!\rm A}}$ elementary operations, where $N_{\!\rm A}$ is the Avogadro's number $\sim {10}^{24}$) on the order of

\begin{equation} \label{6} 
      \frac{1\ {\rm year}}{2^{{10}^{24}}}
      \approx
     \frac{3\cdot {10}^7\ {\rm seconds}}{{10}^{3\cdot {10}^{23}}}\ 
     \sim {10}^{-3\cdot {10}^{23}}\ {\rm seconds}
\;\;\;\;  ,
\end{equation}
\smallskip

\noindent which is by a considerable margin less than one Planck time roughly equal to ${10}^{-43}$ seconds. But within the framework of the laws of physics, for times less than one Planck time apart one can neither measure nor detect any change. Hence, even if there was a device able to do an elementary computing operation in a time as short as ratio (\ref{6}), one would be able to neither measure the result of this operation nor simply detect that this device would have done something. Moreover, for ratio (\ref{6}) there would not be much of a difference between running times of one year and, say, of one hundred billion years (${\sim 10}^{18}$ seconds). So, unless the laws of physics (as we understand them today) were wrong (and consequently some physical processes of a uniquely new nature were possible that allowed construction of übercomputers -- a sort of extraordinary, superior computing devices), no computer would ever be able to execute ${\rm 2}^{N_{\!\rm A}}$ operations in any reasonable amount time.\\

\noindent Hence, in the case, in which the Exponential Time Hypothesis would be true (that would entail the conjecture $\bf{P}\!\!\neq\!\bf{NP}$ is true), the deterministic quantum model of a macroscopic system (built around the exact solutions to the system Schrödinger equation) would be without predictive content inasmuch as there would be no practical means to extract the prediction about the system's future exact state from the Schrödinger equation. In this manner, a Schrödinger cat state -- as a linear combination of the exact (and orthogonalized) solutions to the macroscopic system's Schrödinger equation -- would be predictively contentless and for this reason unavailable for inspection.\\

\section{How the Born rule is explained by computational complexity theory}

\noindent By contrast, a stochastic quantum model of a macroscopic system that is built around inexact (i.e. with a degree of uncertainty) solutions to the system Schrödinger equation might have predictive content even if the Exponential Time Hypothesis was true.\\

\noindent In fact, with a brute-force algorithm it is possible to reach the solution to the Schrödinger equation in reasonable time but only if the state space of a system is limited (as in the case of a system consisting of a few microscopic particles completely isolated from the environment) or when there are system-specific heuristics that can be used to reduce the set of all possible candidate solutions to a limited size.\\

\noindent Take, for example, a composite system comprised of two interacting systems -- a microscopic particle (``test-particle''), whose states are controlled or measured, and a macroscopic system (``environment''), whose constituent microscopic particle states are uncontrolled and unmeasured. As the environment microscopic states are ignored, the set of all possible candidate solutions to the Schrödinger equation for the given composite system can \textit{effectively} (i.e., for all practical purposes) be reduced to the set of the test-particle candidate solutions. In doing so, one would get an inexact yet fast (and so feasible) solution describing (in probability terms) the test-particle interacting with the environment.\\

\noindent Let us show in detail how this heuristics works. At the time $T_{\! f}=T_{\! i}+\tau , \tau \ge 0$ the state of the composite system ``particle + environment'' (whose Hilbert space is the tensor product ${\mathcal H}={{\mathcal H}}_p\otimes {{\mathcal H}}_{\varepsilon }$ of the two Hilbert spaces -- ${{\mathcal H}}_p$ of the particle and ${{\mathcal H}}_{\varepsilon }$ of the environment) is related to that at the initial time $T_{\! i}$ by the following deterministic relation:

\begin{equation} \label{7} 
   \!\left.\left|{\Psi }_{T_{\! i}+\tau }\!\right.\right\rangle
   =
   U\!\left(T_{\! i}+\tau ,T_{\! i}\right)\!\left.\left|{\Psi }_{T_{\! i}}\!\right.\right\rangle
   =
   {\exp \!\left(-\frac{i\tau }{\hbar }H\!\right)\ }\!\!\!\left.\left|{\Psi }_{T_{\! i}}\!\right.\right\rangle
\;\;\;\;  .
\end{equation}
\smallskip

\noindent In this relation, the initial state-vector of the composite system $\!\left.\left|{\Psi }_{T_{\! i}}\!\right.\right\rangle$ is the direct product of the state-vector $\!\left.\left|{\psi}_{T_{\! i}}\!\right.\right\rangle$ in ${{\mathcal H}}_p$ and the state-vector $\left.\left|{\varepsilon }_{T_{\! i}}\!\right.\right\rangle$ in ${{\mathcal H}}_{\varepsilon }$

\begin{equation} \label{8} 
   \!\left.\left|{\Psi }_{T_{\! i}}\!\right.\right\rangle
   =
   \!\left.\left|{\psi}_{T_{\! i}}\!\right.\right\rangle\!\otimes\!\!\left.\left|{\varepsilon }_{T_{\! i}}\!\right.\right\rangle
   =
  \sum_j{c_j\!\left.\left|{\psi }_j\!\right.\right\rangle}\!\otimes\!\sum_k{{\alpha }_k\!\left.\left|{\varepsilon }_k\!\right.\right\rangle}
\;\;\;\;  ,
\end{equation}
\smallskip

\noindent where the orthonormal basis vectors $\!\left.\left|{\psi }_j\!\right.\right\rangle$ and $\!\left.\left|{\varepsilon }_k\!\right.\right\rangle$ spanning the spaces ${{\mathcal H}}_p$ and ${{\mathcal H}}_{\varepsilon }$ are the exact solutions to the Schrödinger equations for the particle Hamiltonian $H_p$

\begin{equation} \label{9} 
   i\hbar\:\displaystyle \frac{\partial }{\partial t}\!\left.\left|{\psi }_j\!\right.\right\rangle
   =
   H_p\!\!\left.\left|{\psi }_j\!\right.\right\rangle
\end{equation}
\smallskip

\noindent and for the environment Hamiltonian $H_{\varepsilon }$

\begin{equation} \label{10} 
   i\hbar\:\displaystyle \frac{\partial }{\partial t}\!\left.\left|{\varepsilon }_k\!\right.\right\rangle
   =
   H_{\varepsilon }\!\!\left.\left|{\varepsilon }_k\!\right.\right\rangle
\;\;\;\;  ,
\end{equation}
\smallskip

\noindent $c_j$ and ${\alpha }_k$ denote complex coefficients of the superpositions, while the Hamiltonian of the composite system $H$ can be presented (at least during interaction time $\tau $) entirely by the interaction term $H_{{\rm int}}$

\begin{equation} \label{11} 
   H
   \cong
   H_{{\rm int}}
   =
   \sum_j{\!\left.\left|{\psi }_j\!\right.\right\rangle\!\left\langle \left.{\psi }_j\right|\right.
   \!\otimes \!
   \sum_k{A_{jk}\!\left.\left|{\varepsilon }_k\!\right.\right\rangle\!\left\langle \left.{\varepsilon }_k\right|\right.}}
\;\;\;\;  ,
\end{equation}
\smallskip

\noindent in which $\!\left.\left|{\psi }_j\!\right.\right\rangle\!\left\langle \left.{\psi }_j\right|\right.$ and $\!\left.\left|{\varepsilon }_k\!\right.\right\rangle\!\left\langle \left.{\varepsilon }_k\right|\right.$ are the operators acting on ${{\mathcal H}}_p$ and ${{\mathcal H}}_{\varepsilon }$, respectively, and $A_{jk}$ stand for the interaction coefficients. So, as it is readily seen from the following expression

\begin{equation} \label{12} 
   \!\left.\left|{\Psi }_{T_{\! i}+\tau }\!\right.\right\rangle
   =
      \!\left(
         I
         -
         \frac{i\tau }{\hbar }
         \sum_j{\!\left.\left|{\psi }_j\!\right.\right\rangle\!\left\langle \left.{\psi }_j\right|\right.
         \!\otimes \!
         \sum_k{A_{jk}\!\left.\left|{\varepsilon }_k\!\right.\right\rangle\!\left\langle \left.{\varepsilon }_k\right|\right.}}
      \!\right)
      \!\sum_j{c_j\!\left.\left|{\psi }_j\!\right.\right\rangle}
      \!\otimes\!
      \sum_k{{\alpha }_k\!\left.\left|{\varepsilon }_k\!\right.\right\rangle}
\;\;\;\;  ,
\end{equation}
\smallskip

\noindent to extract information about the state of the composite system $\!\left.\left|{\Psi }_{T_{\! i}+\tau }\!\right.\right\rangle$ at some moment $T_{\! i}+\tau $ one has to know the basis vectors $\!\left.\left|{\psi }_j\!\right.\right\rangle$ and $\!\left.\left|{\varepsilon }_k\!\right.\right\rangle$, but to obtain them the Schrödinger equations (\ref{9}) and (\ref{10}) must be solved, of course.\\

\noindent 

\noindent The equation (\ref{9}) can assuredly be solved by brute-force search in reasonable time (due to the limited dimensionality of the test-particle state space ${{\mathcal H}}_p$, which before the interaction may be considered as completely isolated from the environment state space ${{\mathcal H}}_{\varepsilon }$), whereas the equation (\ref{10}) cannot. Therefore, to obtain information about the state $\!\left.\left|{\Psi }_{T_{\! i}+\tau }\!\right.\right\rangle$ in the case, in which the Exponential Time Hypothesis holds, we will allow uncertainties in the interaction coefficients $A_{\!jk}$ associated with different microscopic configurations of the environment -- arrangements of its microscopic constituent particles (in view of the fact that those particles are uncontrolled and unmeasured) such that

\begin{equation} \label{13} 
   \forall j,k:
      \ \ \ \
      A_{jk}
      =
      \widetilde{A_j}
      +
      a_{jk}\!\left(\!\omega \!\right)
\;\;\;\;  ,
\end{equation}
\smallskip

\noindent where $\widetilde{A_j}$ are estimates for the interaction coefficients, which can be taken as roughly proportional to the number of electrons in the environment (given that the interaction between the test-particle and its environment can be assumed to be due to the Coulomb force), and $a_{jk}\!\left(\!\omega \!\right)$ are real-valued random variables of equal (among different environment microstates) distribution

\begin{equation} \label{14} 
   \forall j,k:
      \ \ \ \
      a_{jk}\!\left(\!\omega \!\right)
      \sim 
      a_j\!\left(\!\omega \!\right)
\end{equation}
\smallskip

\noindent defined on a set of possible outcomes, the sample space $\Omega $, as

\begin{equation} \label{15} 
   \forall j:
      \ \ \ \
      \left\{ 
             \omega \in \Omega :
             \ \ \
             \left|
             a_j\!\left(\!\omega \!\right)
             \right|
             \le
             \widetilde{A_j}
     \right\}
\;\;\;\;  .
\end{equation}
\smallskip

\noindent Introduced in this manner uncertainties will effectively convert the operator $\sum_k{A_{jk}\!\left.\left|{\varepsilon }_k\!\right.\right\rangle\!\left\langle \left.{\varepsilon }_k\right|\right.}$ (which acts on the environment state space ${{\mathcal H}}_{\varepsilon }$) into the product of a stochastic scalar and the unit operator $\sum_k\!\!\left.\left|{\varepsilon }_k\!\right.\right\rangle\!\left\langle \left.{\varepsilon}_k\right|\right.\!=\!\hat{1}$

\begin{equation} \label{16} 
   \forall j:
      \ \ \ \
      \sum_k{A_{jk}\!\left.\left|{\varepsilon }_k\!\right.\right\rangle\!\left\langle \left.{\varepsilon }_k\right|\right.} 
      \sim 
      \left(
         \widetilde{A_j}
         +
         a_j\!\left(\!\omega \!\right)
      \right)
      \hat{1}
\;\;\;\;  .
\end{equation}
\smallskip

\noindent In turn, the resulted equalities (\ref{16}) will transform the deterministic expression (\ref{11}) for the interaction Hamiltonian into a stochastic one

\begin{equation} \label{17} 
   H_{{\rm int}}\!\left(\!\omega \!\right)
   \sim
   \sum_j{
      \left.\left(
         \widetilde{A_j}
         +
         a_j\!\left(\!\omega \!\right)
      \!\right.\right)
      \!\!\left.\left|{\psi }_j\!\right.\right\rangle\!\left\langle \left.{\psi }_j\right|\right.
      \!\!\otimes
      \!\hat{1}
      }
\end{equation}
\smallskip

\noindent and in this way will preclude the necessity of solving the environmental Schrödinger equation (\ref{10}) to obtain the incomplete (as the environmental microstates $\left.\left|{\varepsilon }_{T_{\! i}}\!\right.\right\rangle$ are unknown) prediction for the final state $\!\left.\left|{\Psi }_{T_{\! i}+\tau }\!\right.\right\rangle$

\begin{equation} \label{18} 
      \!\left.\left|{\Psi }_{T_{\! i}+\tau }\!\right.\right\rangle
      =
      \!\left(
         I
         -
         \frac{i\tau }{\hbar }H_{{\rm int}}\!\left(\!\omega \!\right)
      \!\right)
      \!\sum_j{c_j\!\left.\left|{\psi }_j\!\right.\right\rangle}
      \!\otimes
      \!\left.\left|{\varepsilon }_{T_{\! i}}\!\right.\right\rangle
      \sim
      \!\left.\left|{\psi }_{T_{\! i}+\tau }\!\left(\!\omega \!\right)\!\right.\right\rangle
      \!\otimes
      \!\left.\left|{\varepsilon }_{T_{\! i}}\!\right.\right\rangle
\;\;\;\;  ,
\end{equation}
\smallskip

\noindent which will, nonetheless, contain information -- albeit inexact one -- about the state of the test-particle at the final time $T_{\! i}+\tau $ after the interaction with the environment:

\begin{equation} \label{19} 
   \left\{
      \omega \in \Omega :
      \ \ \
      \!\left.\left|{\psi }_{T_{\! i}+\tau }\!\left(\!\omega \!\right)\!\right.\right\rangle
      =
      \!\sum_j{c_j\!\left.\left|{\psi }_j\!\right.\right\rangle}
      \,{\exp
         \!\left(
            -
            \frac{
               \widetilde{A_j}
               +
               a_j\!\left(\!\omega \!\right)
            }
            {\hbar }
            \,{i\tau }
         \!\right)\ 
      }
   \!\!\!\right\}
\;\;\;\;  .
\end{equation}
\smallskip

\noindent As follows, the random state-vector $\!\left.\left|{\psi }_{T_{\! i}+\tau }\!\left(\!\omega \!\right)\!\right.\right\rangle$ does not represent a single, fixed final state of the test-particle; rather it takes on a set of possible different final states. That is to say, the vector $\!\left.\left|{\psi }_{T_{\! i}+\tau }\!\left(\!\omega \!\right)\!\right.\right\rangle$ associates states of the test-particle at the final time $T_{\! i}+\tau $ with instances $\omega $ of a yet-to-be-performed experiment, so that $\!\left.\left|{\psi }_{T_{\! i}+\tau }\!\left(\!\omega \!\right)\!\right.\right\rangle$ will vary from instance to instance as the experiment is repeated.\\

\noindent 

\noindent This means that there must be a probability distribution associated with the random state-vector $\!\left.\left|{\psi }_{T_{\! i}+\tau }\!\left(\!\omega \!\right)\!\right.\right\rangle$ that allows the computation of the probabilities of the possible final states. But in accordance with the postulates of quantum mechanics, the state-vector of the test-particle $\!\left.\left|{\psi }_{T_{\! i}+\tau }\!\left(\!\omega \!\right)\!\right.\right\rangle$ determines everything that can be known about this test-particle. It can be inferred from here that the probability distribution associated with the state-vector $\!\left.\left|{\psi }_{T_{\! i}+\tau }\!\left(\!\omega \!\right)\!\right.\right\rangle$ must be determined by the vector $\!\left.\left|{\psi }_{T_{\! i}+\tau }\!\left(\!\omega \!\right)\!\right.\right\rangle$ itself (otherwise, quantum theory cannot be considered complete).\\

\noindent 

\noindent Yet, mathematically, using the complex vector $\!\left.\left|{\psi }_{T_{\! i}+\tau }\!\left(\!\omega \!\right)\!\right.\right\rangle$ is impossible to define a probability measure -- a real-valued non-negative function that must return results in the unit interval $\left[0,1\right]$ (producing 0 for the empty set and 1 for the entire sample set $\Omega $) and satisfy the countable additivity property. On the other hand, according to the Gleason's theorem \cite{Gleason,Caves,Schlosshauer05}, if one would like to assign a probability measure to the vector $\!\left.\left|{\psi }_{T_{\! i}+\tau }\!\left(\!\omega \!\right)\!\right.\right\rangle$, the only possible choice is ${\left|\left\langle \varphi \!\!\right.\left|\left.{\psi }_{T_{\! i}+\tau }\!\left(\!\omega \!\right)\right\rangle \!\right.\right|}^2$, the modulus squared of the scalar product of $\!\left.\left|{\psi }_{T_{\! i}+\tau }\!\left(\!\omega \!\right)\!\right.\right\rangle$ and some arbitrary but fixed vector $\left|\left.\varphi \right\rangle \right.$. Choosing the initial state-vector $\!\left.\left|{\psi }_{T_{\! i}}\!\right.\right\rangle$ of the test-particle as such a fixed vector, one will have the following probability measure turning the sample space $\Omega $ into a probability space:

\begin{equation} \label{20} 
   \left\{
      \omega \in \Omega :
      \ \ \
      {\left|\left\langle {\psi }_{T_{\! i}} \!\!\right.\left|\left.{\psi }_{T_{\! i}+\tau }\!\left(\!\omega \!\right)\right\rangle \!\right.\right|}^2
   \right\}
   =
   \left[0,1\right]
\;\;\;\;  ,
\end{equation}
\smallskip

\noindent where

\begin{equation} \label{21} 
      {\left|\left\langle {\psi }_{T_{\! i}} \!\!\right.\left|\left.{\psi }_{T_{\! i}+\tau }\!\left(\!\omega \!\right)\right\rangle \!\right.\right|}^2
      =
      {\left|
         \!\left(
            \!\sum_g{{c^*}_{\!\!g}}\left\langle \left.{\psi }_g\right|\right.
         \!\right)
         \!\!\left(
            \!\sum_j{c_j\!\left.\left|{\psi }_j\!\right.\right\rangle}
            \,{\exp
              \!\!\left(
                  -
                  \frac{
                     \widetilde{A_j}
                     +
                     a_j\left(\!\omega \!\right)
                  }
                  {\hbar }
                  \,{i\tau }
              \!\right)
              }
         \!\!\right)
      \!\right|}^2
\;\;\;\;  ,
\end{equation}
\smallskip

\noindent provided that $\sum_j{{\left|c_j\right|}^2}\!=\!1$. Performing the experiment many times, one can find a typical (of this experiment) value for the probability of transitioning the test-particle from the initial state $\!\left.\left|{\psi }_{T_{\! i}}\!\right.\right\rangle$ to the final state $\!\left.\left|{\psi }_{T_{\! i}+\tau }\right.\!\right\rangle$ by averaging ${\left|\left\langle {\psi }_{T_{\! i}} \!\!\right.\left|\left.{\psi }_{T_{\! i}+\tau }\!\left(\!\omega \!\right)\right\rangle \!\right.\right|}^2$ over the entire sample set $\Omega $:

\begin{equation} \label{22} 
      P\left(
         \!\left.\left|{\psi }_{T_{\! i}}\!\right.\right\rangle
         \!\to
         \!\left.\left|{\psi }_{T_{\! i}+\tau }\right.\!\right\rangle
      \right)
      =
      \overline{
      {\left|\left\langle {\psi }_{T_{\! i}} \!\!\right.\left|\left.{\psi }_{T_{\! i}+\tau }\!\left(\!\omega \!\right)\right\rangle \!\right.\right|}^2
      }
\;\;\;\;  .
\end{equation}
\smallskip

\noindent Assuming that the state space of the test-particle is the $n$-dimensional complex Hilbert space $C^{\,n}$ (where $n$ is limited) and for the sake of simplicity supposing that all ${\left|c_j\right|}^2\!\!=\!\frac{1}{n}$, one can find from Eq.(\ref{21})

\begin{equation} \label{23}
   {\left|\left\langle {\psi }_{T_{\! i}} \!\!\right.\left|\left.{\psi }_{T_{\! i}+\tau }\!\left(\!\omega \!\right)\right\rangle \!\right.\right|}^2
   =
   \frac{1}{n}
   +
   \frac{2}{n^2}
   \sum^{n-1}_{j=1}{
      \sum^n_{g=j+1}{
         \!\!\left( \cos {\Xi}_{jg} \,\cos {\xi}_{jg}\!\left(\!\omega\!\right) - \sin {\Xi}_{jg} \,\sin {\xi}_{jg}\!\left(\!\omega\!\right) \right) }
   }
\;\;\;\;  ,
\end{equation}
\smallskip

\noindent where the angles ${\Xi}_{jg}$ are

\begin{equation} \label{24}
   {\Xi}_{jg}
   =
   \frac{\widetilde{A_j}-\widetilde{A_g}}{\hbar }\,\tau
\;\;\;\;  ,
\end{equation}
\smallskip

\noindent and the random angles ${\xi }_{jg}\!\left(\!\omega \!\right)$ are defined as

\begin{equation} \label{25}
   {\xi}_{jg}\!\left(\!\omega \!\right)
   =
   \frac{a_j\!\left(\!\omega \!\right)-a_g\!\left(\!\omega \!\right)}{\hbar }\,\tau
\;\;\;\;  .
\end{equation}
\smallskip

\noindent The total span of the random angles can be assessed by their maximum and minimum limits:

\begin{equation} \label{26}
   {\mathop{\max }_{\omega \in \Omega } {\xi }_{jg}\!\left(\!\omega \!\right)\ }
   \!{=}
   -{\mathop{\min }_{\omega \in \Omega } {\xi }_{jg}\!\left(\!\omega \!\right)\ }
   \!\approx
   \frac{\widetilde{A_j}+\widetilde{A_g}}{\hbar }\,\tau
\;\;\;\;  ;
\end{equation}
\smallskip

\noindent so, assuming that the random angles ${\xi }_{jg}\!\left(\!\omega \!\right)$ are spread uniformly within these limits (because the uniform distribution is the one that makes the least claim to being informed about the interaction coefficients $A_{jk}$ associated with uncontrolled and unmeasured microscopic configurations of the environment beyond knowing the approximate limits of $A_{jk}$), one can find the average values of the functions of ${\xi }_{jg}\!\left(\!\omega \!\right)$ over the sample set $\Omega$:

\begin{equation} \label{27}
   \overline{
      {\cos {\xi }_{jg}\!\left(\!\omega \!\right)\ }
   }
   \!\approx 
   \frac{\hbar }{\left(\widetilde{A_j}+\widetilde{A_g}\right)\tau }
   {\,\sin  \frac{\widetilde{A_j}+\widetilde{A_g}}{\hbar }\,\tau \ }
\;\;\;\;  ,
\end{equation}
\smallskip

\begin{equation} \label{28}
   \overline{{\sin  {\xi }_{jg}\!\left(\!\omega \!\right)\ }}
   =
   0
\;\;\;\;  .
\end{equation}
\smallskip

\noindent To define the number of electrons in the environment one can assume that only those environmental electrons that are within a radius of $R=cT$ (where $c$ is the speed of light, $T$ is the time allotted for the interaction, which is typically of order ${10}^{-3}$ seconds) can influence the test-particle. It is obvious that in the normal environment the number of electrons within the causal horizon $R$ (and thus the estimates $\widetilde{A_j}$ and $\widetilde{A_g}$) will be of a considerable magnitude. This means that after a very short period of the interaction, the argument of the sine function in (\ref{27}) will be close to infinity and consequently the whole right hand side of (\ref{27}) will be close to zero. So, if the state of the test-particle is initially given by the superposition state $\left.\left|{\psi }_{T_{\! i}}\!\right.\right\rangle=\sum^n_j{\frac{1}{\sqrt{n}}\left.\left|{\psi }_j\!\right.\right\rangle}$, then at the time $T_{\! i}+\tau $ the transition probability (\ref{22}) will be equal to that following from the Born rule

\begin{equation} \label{29}
      P\left(
         \!\left.\left|{\psi }_{T_{\! i}}\!\right.\right\rangle
         \!\to
         \!\left.\left|{\psi }_{T_{\! i}+\tau }\right.\!\right\rangle
      \right)
   \approx 
   P\left(
      \!\left.\left|{\psi }_{T_{\! i}}\!\right.\right\rangle
      \!\to
      \!\left.\left|{\psi }_j\right.\!\right\rangle
   \right)
   =
   \frac{1}{n}
\end{equation}
\smallskip

\noindent meaning that upon the interaction with the environment the initial state $\!\left.\left|{\psi }_{T_{\! i}}\!\right.\right\rangle$ will `collapse' in the sense that $\!\left.\left|{\psi }_{T_{\! i}}\!\right.\right\rangle$ will change to $\!\left.\left|{\psi }_j\right.\!\right\rangle$.\\

\noindent In this way, the reduction postulate and the Born rule can be considered as a mere shortcut, a way to get the last result without using the presented above heuristic.\\

\section{Concluding remarks}

\noindent Since the earliest years of quantum theory, it has become increasingly evident that the rapid rate, at which the Schrödinger equation grows to be more complicated as the size of a system increases, makes the task of deriving complete properties of macroscopic objects from their microstates simply hopeless. Yet, such a problem was never considered as something fundamental since one may always hope that the Schrödinger equation will certainly be solved someday at least numerically, because numerical solutions are always reachable if only enough computational resources are thrown at them.\\

\noindent However, in all likelihood, such is not the actual state of things in our real physical world. As it was shown in this paper, given an exact generic algorithm capable of solving exactly the Schrödinger equation for an arbitrary physical Hamiltonian (that is, for any and all possible physical systems), any problem in the \textbf{NP} complexity class can be solved with only polynomially more work. This implies that unless the Exponential Time Hypothesis fails (which will be highly surprising), coming up with the exact solution to Schrödinger's equation for an arbitrary system will inevitably involve using an algorithm which is not significantly faster than brute force search over an exponential in size set of all possible candidate solutions. As a result, computational resources required by this algorithm will grow so rapidly with the system microscopic constituent particle number that bringing any additional resources to bear on the algorithm will be just of no value. And so, for anyone living in the real physical world (of limited computational recourses) the Schrödinger equation will turn out to be simply unsolvable for macroscopic objects and accordingly inapplicable to their time evolution portrayal.\\

\noindent In other words, in the case, in which the Exponential Time Hypothesis holds, it is impossible to overlap deterministic quantum and classical descriptions in order to obtain a rigorous derivation of classical properties from quantum mechanics.\\

\noindent As said, another foundational problem in quantum theory is how to reconcile the linear, deterministic evolution described by the Schrödinger equation with the occurrence of random, definite measurement outcomes. In this paper, randomness (and associated with it probability) is entered as a way to obtain the prediction about the final state of the test-particle interacting with the environment without solving the environmental Schrödinger equation -- an intractable computational problem. For to allow statistical uncertainty in the description of a system is effectively equivalent to making the description less detailed, which in turn reduces the number of possible candidate solutions, needed to search over to find the correct one. Clearly, had the Schrödinger equation have the efficient generic algorithm that could solve it exactly for all possible physical Hamiltonians in polynomial time, the predicted state of the composite system ``particle + environment'' would be given by a deterministic expression deprived of any randomness.\\

\end{document}